\documentclass[sigconf]{acmart}
\usepackage{changepage}
\usepackage{subfigure}

\AtBeginDocument{%
  \providecommand\BibTeX{{%
    \normalfont B\kern-0.5em{\scshape i\kern-0.25em b}\kern-0.8em\TeX}}}

 \setcopyright{none}

\newtheorem{resques}{Research Question}
\begin{document}

\title{Large Scale Analysis of Multitasking Behavior \\During Remote Meetings}

\author{Hancheng Cao}
\authornote{The work was done when Hancheng Cao interned at Microsoft.}

\affiliation{%
  \institution{Stanford University}
  \city{Stanford}
  \state{CA}
  \country{USA}
}
\email{hanchcao@stanford.edu}

\author{Chia-Jung Lee}
\affiliation{%
  \institution{Amazon}
  \city{Seattle}
  \state{WA}
  \country{USA}
}
\email{cjlee@amazon.com}
\authornote{The work was done when Chia-Jung Lee worked full time at Microsoft.}

\author{Shamsi Iqbal}
\affiliation{%
  \institution{Microsoft}
  \city{Redmond}
  \state{WA}
  \country{USA}
}
\email{shamsi@microsoft.com}

\author{Mary Czerwinski}
\affiliation{%
  \institution{Microsoft}
  \city{Redmond}
  \state{WA}
  \country{USA}
}
\email{marycz@microsoft.com}

\author{Priscilla Wong}
\affiliation{%
  \institution{University College London}
  \city{London}
  \country{UK}
}
\email{ngoi.wong.13@ucl.ac.uk}

\author{Sean Rintel}
\affiliation{%
  \institution{Microsoft}
  \city{Cambridge}
  \country{UK}
}
\email{serintel@microsoft.com}

\author{Brent Hecht}
\affiliation{%
  \institution{Microsoft}
  \city{Redmond}
  \state{WA}
  \country{USA}
}
\email{Brent.Hecht@microsoft.com}

\author{Jaime Teevan}
\affiliation{%
  \institution{Microsoft}
  \city{Redmond}
  \state{WA}
  \country{USA}
}
\email{teevan@microsoft.com}

\author{Longqi Yang}
\affiliation{%
  \institution{Microsoft}
  \city{Redmond}
  \state{WA}
  \country{USA}
}
\email{Longqi.Yang@microsoft.com}

\renewcommand{\shortauthors}{Cao et al.}

\begin{abstract}

Virtual meetings are critical for remote work because of the need for synchronous collaboration in the absence of in-person interactions. In-meeting multitasking is closely linked to people's productivity and wellbeing. However, we currently have limited understanding of multitasking in remote meetings and its potential impact. In this paper, we present what we believe is the most comprehensive study of remote meeting multitasking behavior through an analysis of a large-scale telemetry dataset collected from February to May 2020 of U.S. Microsoft employees and a 715-person diary study. Our results demonstrate that intrinsic meeting characteristics such as size, length, time, and type, significantly correlate with the extent to which people multitask, and multitasking can lead to both positive and negative outcomes. Our findings suggest important best-practice guidelines for remote meetings (e.g., avoid important meetings in the morning) and design implications for productivity tools (e.g., support positive remote multitasking).

\end{abstract}




\maketitle

\section{Introduction}
Remote work is an essential mode of work across industries~\cite{bloom2015does}. Before COVID-19, many companies had already experimented with or implemented various forms of remote work~\cite{brynjolfsson2020covid}, e.g., recruiting remote employees or allowing employees to work from home part-time. During the pandemic,  34.1\% of Americans
switched to working from home~\cite{brynjolfsson2020covid}, and it is estimated that 37\% of jobs in United States can be done remotely~\cite{dingel2020many}. In fact, many large technology companies, such as Quora and Twitter, have announced that they will allow employees to work from home indefinitely~\cite{adam_dangelo_remote_nodate, jennifer_christie_keeping_nodate}. 

Central to remote work are remote meetings \cite{olson2000distance, cao2020my}, most commonly experienced through video conferencing tools (e.g., Zoom, Microsoft Teams, Google Meet), which help remote team members stay connected, collaborate and function as an organization, despite physical distance. Therefore, it is critical to understand factors that are associated with remote meeting experiences to better support productive and engaging remote collaborations.


In this paper, we focus on one fundamental experience of remote meetings --- in-meeting multitasking. Information work is often governed by multiple tasks and activities that an individual must remember to perform, often in parallel or in rapid succession, a practice that is called \emph{multitasking}~\cite{Rubinstein2001, Wickens2008, czerwinski2004diary}. 
Multitasking behavior is vital, as it is closely linked to one's productivity and well-being~\cite{czerwinski2004diary,madore2020memory} --- increasing the numbers of items to be remembered can wreak havoc with task resumption~\cite{altmannResumption}, attentional focus ~\cite{iqbal2007disruption, mark2004} and prospective memory~\cite{brandimonteProspectiveMemory}. Multitasking during meetings can additionally affect other people and their productivity \cite{iqbal_meeting}.
Prior work has investigated how people engage in multitasking during collaborative activities such as meetings and video chats, both in-person and online \cite{iqbal_meeting, suh-meeting, marlow-meeting, Hembrooke2003TheLA}. 
However, these works are mainly based on small-scale qualitative studies, and to the best of our knowledge, no research to date has reported systematic and comprehensive evidence from large populations. As a result, there is little statistical evidence on the meeting and personal context that are associated with people's propensity to multitask, the activities remote attendees engage in while multitasking, and how such remote multitasking behavior may affect workers and groups. 

Here, we adopt a mixed-methods approach to systematically understand the context, activities, and consequences of multitasking during remote meetings. Specifically, we analyzed a large-scale dataset collected from from February to May 2020 of U.S. Microsoft employees. The dataset contains anonymized events from major productivity tools: Microsoft Teams for remote meetings, Microsoft Outlook for email services, and OneDrive and Sharepoint for accessing and editing files remotely. Multitasking can involve activities that are unrelated to the meeting (e.g., email communications) and related (e.g., notes taking)~\cite{iqbal_meeting}. Therefore we measured emails sent and files edited during remote meetings as a proxy for multitasking and studied the relationship between multitasking and meeting characteristics through controlled regression analysis. Furthermore, we contextualize the evidence from log analysis with verbatims from a 715-person diary study of Microsoft employees globally, run from mid-April to mid August 2020, exploring their remote meeting experiences during COVID-19. 

Our results show that: 1) Multitasking is a common behavior in remote meetings with about 30\% of meetings involving email multitasking. People also reported that multitasking becomes more frequent as meetings move to remote, 2) Intrinsic meeting characteristics, such as meeting size, length, type, and timing, significantly correlate with the extent to which people multitask, e.g., people are more likely to multitask in recurring meetings than in ad hoc meetings, and 3) In-meeting multitasking during remote meetings can lead to both positive (e.g., improve productivity) and negative (e.g., loss of attention)  experiences. Our analysis suggests practical ways people can improve remote meeting experiences. For example,  to reduce unnecessary scheduled and recurring meetings, keep meetings short, avoid intensive meetings early in the morning, and allow space for positive multitasking. Our work implies that productivity tools can help people better manage their in-meeting attention and decide which meetings or parts of the meetings to attend.

The contributions of this paper are threefold: 

\begin{itemize}
\item{We present the first large-scale, empirical study of multitasking behavior during remote meetings, accompanied by rich qualitative evidence. This allows us to understand the factors that correlate with remote meeting multitasking and characterize the motivations and potential consequences of such behavior.} 
\item{We discover several key issues in current remote meeting configurations and suggest actionable guidelines for how people may schedule effective remote meetings (Section~\ref{sec:best_practice}).} 
\item{Our work points to several concrete design implications for future remote collaboration tools, e.g., support meeting ``focus mode'' and positive multitasking.}

\end{itemize}
\section{Background and Related Work}

Our work is built upon and contributes to the rich HCI literature on remote work and multitasking behavior and its manifestations in meetings. Prior research mostly focused on small-scale in-person studies, whereas our work analyzes a large-scale log dataset accompanied by a large diary study, to systematically reveal multitasking patterns during all-remote meetings.

\subsection{Remote Work}
Remote work has long been an important topic of research across scientific fields. Through a Working-From-Home (WFH) experiment at Ctrip, a 16,000-employee Chinese travel agency, Bloom et al.~\cite{bloom2015does} studied the costs and benefits of remote work, where they showed that WFH led to improved performance but reduced promotion rate. Prior work also investigated other aspects of remote work, such as management of workers~\cite{delle2018missing}, organizational design~\cite{valentine2017flash, retelny2014expert}, communication challenges~\cite{he2014qualitative}, team performance~\cite{lix2020timing}, well-being~\cite{cao2020my,whiting2019did}, and emerging roles~\cite{chen2020understanding} and experience~\cite{cao2020your, cao2020you}. Remote work becomes more ubiquitous after the COVID-19 pandemic~\cite{eriksson2020remote, yang2020work}, and remote meetings have become a central place where people stay connected and collaborate with others~\cite{olson2000distance} --- both DeFilippis et al.~\cite{defilippis2020collaborating} and Yang et al.~\cite{yang2020work} demonstrated a significant increase in remote meetings during the pandemic. Here we contribute to the rich literature of remote work by focusing on how in-meeting multitasking, an artifact of traditional in-person meetings, manifested in all-remote settings.

\subsection{Multitasking in the Workplace}
A large body of prior work has focused on how multitasking impacts attention in the workplace, primarily focusing on the distraction caused to an ongoing task that is interrupted by another activity. Czerwinski et al.~\cite{czerwinski2004diary} employed a diary study to show how information workers switch activities due to interruptions in the workplace, focusing on the difficulty of the continuous switching of context. Iqbal and Horvitz~\cite{iqbal2007disruption} studied how external interruptions cause information workers to enter into a ``chain of distraction'' where stages of preparation, diversion, resumption and recovery can describe the time away from an ongoing task. Gonzalez and Mark~\cite{gonzalez2004constant} reported on how information workers conceptualize and organize basic units of tasks and how switching occurs across these conceptual units --- people were found to spend about 12 minutes in one working unit before switching to another. In the mobile domain, Karlson et al.~\cite{karlson2010mobile} and Park et al.~\cite{park2014analysis} found that tasks on mobile phones become fragmented across devices and they identified challenges that exist in resuming these tasks. While most studies have looked at multitasking results from distractions, there is a gap in prior work that characterizes interactions corresponding to multitasking during remote meetings, as discussed in Section~\ref{sec:mutitasking_meeting}. 

\subsection{Factors Associated with Multitasking}

External~\cite{oconaill-frohlich-externalinterruptions, iqbal_meeting, mark-gonzalez-2005, mark-wang-2014-collegestress} and internal~\cite{dabbish-mark-2011, alder2013} interruptions are considered to be the most direct reasons behind multitasking; however there are other indirect factors that are associated with multitasking as well. Past work has shown that personality and organizational environment~\cite{dabbish-mark-2011} are associated with multitasking~\cite{mark-neurotics}. Additionally, previous work~\cite{mark2014bored} has shown that the time of day is associated with workers' focus. On average, people were most focused in their work late-morning (11 a.m.) and mid-afternoon (2-3 p.m.), which is known as the ``double peak day'' of information workers' rhythm of work. 
Day of the week also play important role in people's attention level and multitasking behavior. Mark et al.~\cite{mark2014bored} found a relationship between online activity and Mondays, the day when people report being the most bored, but at the same time also the most focused. However, most of the prior studies are focused on in-person work settings, and it is unclear to what extent these patterns are applicable to remote meetings. Also, the existing evidence is mainly based on small-scale data that may not generalize to large organizations. Our study confirms the associations between time, distractions and multitasking in remote meetings through large-scale log analysis and greatly extends prior knowledge by investigating a broader range of meeting characteristics, such as meeting size, type, and length.

\subsection{Multitasking during Meetings}
\label{sec:mutitasking_meeting}

While multitasking during one's own work mostly impacts personal productivity, special consideration of multitasking during meetings is warranted, as this can additionally impact other colleagues and their productivity~\cite{iqbal_meeting}. Past work has looked at how people engage in multitasking both during in-person meetings and presentations~\cite{iqbal_meeting, Hembrooke2003TheLA, barkhuus-laptop}, as well as online collaborative activities, such as remote meetings leveraging subjective feedback or perceptions~\cite{marlow-meeting, suh-meeting}. For example, in educational settings, the use of laptops during a lecture has been shown to have a negative impact on attention, where students tend to engage in activities such as web-surfing or emailing rather than activities related to the lecture~\cite{barkhuus-laptop, Hembrooke2003TheLA}. However, in other studies, people who multitask during in-person meetings report to do so in order to interleave other important activities as they peripherally pay attention to the meeting and engage only when relevant~\cite{iqbal_meeting}. In online settings, both meeting related and personal multitasking are seen as ways people's attention could divert from the actual conversation, though multitasking on a single screen is considered more acceptable than when multitasking is happening on a different screen - often presumed to be unrelated to the meeting~\cite{marlow-meeting}. Similarly, a study on video-chats among teens found that boredom was the main reason why teens would multitask during a video chat, wherein they would engage in scrolling social feeds or play games~\cite{suh-meeting}. 

While prior work on meetings and multitasking focus primarily subjective perceptions, no research to the best of our knowledge has looked at large-scale log data accompanied by qualitative evidence on what people are engaging in while attending a meeting and under what conditions people tend to engage in when multitasking. Our analysis of actual interactions can complement subjective perceptions around multitasking motivations, behaviors and potential consequences, and can provide insights into how to conduct meetings for maximal effectiveness. 

\section{Method}

To systematically understand multitasking patterns during remote meetings, we proposed the following research questions to guide our research.

\begin{resques}[RQ\ref{resques:volume}] \label{resques:volume}
 How much multitasking is happening during remote meetings? 
\end{resques}

\begin{resques}[RQ\ref{resques:when}] \label{resques:when}
 What factors are associated with multitasking during remote meetings? 
\end{resques}

\begin{resques}[RQ\ref{resques:why}] \label{resques:why}
 What do people do when multitasking during remote meetings? 
\end{resques}

\begin{resques}[RQ\ref{resques:consequence}] \label{resques:consequence}
 What are the consequences of multitasking during remote meetings?
\end{resques}

As noted above, we adopt a mixed-methods approach to address these questions: analysis of a large-scale anonymized telemetry dataset coupled with a diary study of people's perceptions and subjective experiences with respect to in-meeting multitasking.

\subsection{Regression Analysis on Large-Scale Telemetry Dataset}
\label{sec:regression_analysis}

\textbf{Data Preparation.} We collected metadata (without any content information) on remote meetings (Microsoft Teams), email usage (Microsoft Outlook), and file edits (Onedrive/Sharepoint) of US employees from Microsoft. The majority of work and communication in Microsoft are carried out through these platforms. We collected four separate week-long snapshots of data from February to May, 2020: 1) February 24-28, which represents a period of pre-COVID, mostly co-located work, 2) March 23-27, which represents the company’s transition phase from mostly co-located work to remote work, and 3) April 20-24 and 4) May 18-22, to represent fully remote work periods. While we leveraged all four periods to study work rhythm and derive statistics on multitasking behavior, our regression analysis focused on the snapshot from May 18-22, when employees were fully settled into working from home.

\begin{figure*}[tb]
      \begin{subfigure}[Distribution of meeting duration]{
        \label{fig:meeting_duration}
        \includegraphics[width=0.3\textwidth]{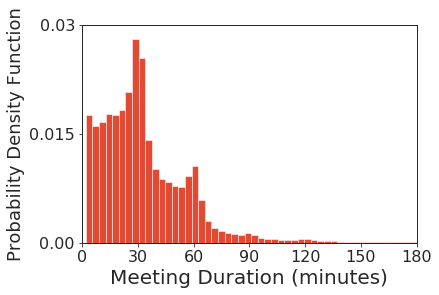}}
       \end{subfigure}
       \begin{subfigure} [Distribution of meeting size]{
        \label{fig:meeting_size}
        \includegraphics[width=0.3\textwidth]{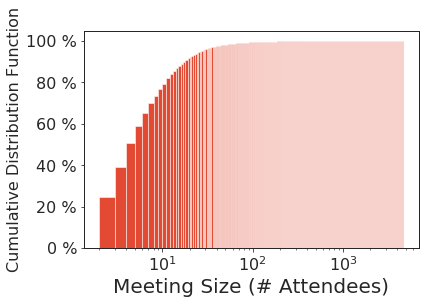}}
       \end{subfigure}
       \begin{subfigure} [Distribution of meeting types]{
        \label{fig:meeting_type}
        \includegraphics[width=0.3\textwidth]{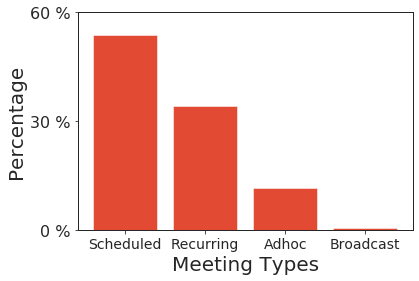}}
       \end{subfigure}
 \caption{Distributions of virtual meeting duration, meeting size, and meeting types measured through our collected telemetry data. We observe two clear peaks surrounding 30 mins and 60 mins in distribution of meeting duration, indicating the popularity of scheduling 30 mins and 60 mins meetings. We observe a long tail distribution of meeting size, where over 20\% of all meetings are one on one meetings. Finally, we observe the majority of meetings are scheduled meetings, followed by recurring meetings, adhoc meetings and broadcast meetings. The data distributions motivate us to discretize features into bins specified in Table 1.}
 \label{fig:workrhythm}
\end{figure*}

Specifically, for each meeting hosted on Microsoft Teams\footnote{We only included meetings that are longer than two minutes to filter out data noise. We also filtered out meetings lasting longer than 3 hours, which is likely due to the fact people forget to leave the meeting.}, we collected the start and end timestamps of the meeting \footnote{Meeting start and end timestamps was logged on a per-person basis, i.e, the exact timestamp that each person attended and left a meeting. The average standard deviation of meeting duration among people joining the same meeting is about 2.1\% of the maximum meeting duration, indicating that the same meeting generally has similar length across participants.}, meeting size\footnote{Meeting size was logged as the number of all people ever connected to a meeting. Due to aggregated nature of telemetry data, it is impossible to measure the exact maximum concurrent participants.}, and the meeting type (ad hoc, scheduled, recurring or broadcast\footnote{Meetings that have more than 250 attendees.}), and their distributions are presented in Fig.~\ref{fig:meeting_duration}. As shown in Table.~\ref{table:features}, we discretized continuous meeting attributes in the following ways to ensure robustness of the regression analysis. We grouped meeting duration into four categories --- 0-20 mins, 20-40 mins, 40-80 mins, and >80 mins because of the popularity of 30 mins and 60 mins meetings (Fig.~\ref{fig:meeting_duration}). For meeting size, due to its long-tail nature, we split it into five roughly equal sized bins. We considered morning, afternoon, and after hours as three categories for hour of the day in order to align it with common work rhythms in Microsoft.\footnote{We tested different bucketing strategies and it produced robust results.} Furthermore, we tested the meeting attribute correlation, where we find rather weak correlations ($\left|r\right|<0.15$) among different attribute groups. For instance, the correlation between meeting time (e.g., morning) and meeting size (e.g., >10 attendees) is around 0.07. The weak dependency between meeting attributes motivated us to directly include them in the regression analysis rather than trying to cluster various meeting types and study those clusters in the regression. Finally, we note that the telemetry data only records virtual meetings, so February data (before pandemic in U.S.) does not reflect total volume of meetings people attended, but all meetings after the pandemic are recorded.

We focused on two multitasking behavior during remote meetings that can be measured through available telemetry data: email multitasking and file edit multitasking. To enable the analysis, on Microsoft Outlook, we collected the time when people actively send, respond to, or forward an email. On file platforms, we recorded events related to editing productivity-related files, including Excel, Powerpoint, Word, and PDFs. Note that due to the nature of the data, we are not able to differentiate whether the files are related to the meeting or not, thus the measured multitasking file behavior might be related to the meeting (e.g. notes taking).

To ensure people's privacy with the telemetry data, we performed pre-processing to de-identify workers before the data was obtained by researchers. Access to the data was strictly limited to authorized members of the research team who went through extensive privacy and ethical training.

\textbf{Regression Analysis.} We joined three data sources by unique user identifiers (anonymized), which resulted in 34,524 (meeting, user) records. For each (meeting, user) pair, we labeled it as email multitasking ($Y=1$) if the user was found to have at least one active email action during the meeting, otherwise a non-multitasking label was assigned ($Y=0$). File multitasking was labeled similarly. We conducted a regression analysis to understand the relationships between multitasking and meeting characteristics, while controlling for individual variances since the individual tendency to multitask may confound the estimation. We leveraged stratification \cite{morgan2015counterfactuals} to group all records by worker and used conditional logistic regression to fit the model. The binary meeting indicators used in the regression model are defined in Table.~\ref{table:features}. To account for the correlations between the records from the same meetings, we grouped standard errors at the meeting level. We present regression results in terms of estimated odds ratio and their statistical significance in Fig.~\ref{fig:Regression_email}. We did not find significant correlations between file-related multitasking and meeting characteristics, which could be attributed to the fact that files edited are often related to the meeting (Section~\ref{sec:file_multitasking_qualitative}). An alternative analysis using Generalized Linear Mixed Effects Models~\cite{seabold2010statsmodels} show qualitatively similar results (Appendix~\ref{sec:supp}).

\begin{table*}[t]
\centering
\vspace{-0.1cm}
\caption{Meeting property and discretized bins used for regression analysis. For each meeting property, we create dummy variables and then left one out as baseline. For instance, Friday is left as baseline for the day of the week meeting property, and all effects are relative to meetings on Friday.} 
\vspace{-0.2cm}
\scalebox{0.8}{
\begin{tabular}{@{\hspace{0.1cm}}lc@{\hspace{0.1cm}}}
\toprule
\textbf{Meeting Property}                               & \textbf{Discretized bins} \\
\midrule
Meeting duration &  0-20 mins meetings (baseline), 20-40 mins meetings, 40-80 mins meetings, >80 mins meetings \\
Meeting size &   2 attendees (baseline), 3 attendees, 4-5 attendees, 6-10 attendes, >10 attendees.\\
Meeting types  &  scheduled, recurring, ad hoc (baseline), broadcast \\
Hour of the day &  morning: 7am - noon, afternoon: noon - 7pm, after hours: 7pm - 7am (baseline)\\
Day of the week &  Monday, Tuesday, Wednesday, Thursday, Friday (baseline) \\

\bottomrule
\end{tabular}
}
\label{table:features}
\vspace{-0.0cm}
\end{table*}


\textbf{Limitations of Telemetry Data.} While the large-scale telemetry data provides us with a lens onto how people behave during meetings, our emphasis on preserving privacy means that we unfortunately do not have access to all behavioral details. The data was collected in anonymized, aggregated form and does not have sensitive attributes that can potentially reveal the identity of an individual employee or a group in a corporation, such as functional labels, job roles, organizational charts or participant social demographics. Email is perceived as stateless communication in the organization we studied; therefore, fine-grained events e.g., email read was not recorded. Similarly, multitasking behavior in 3rd party platforms and non-digital spaces (e.g., house chores), is not measured due to lack of instrumentation. Finally, telemetry data lacks information on the reasons for and consequences of multitasking. We strive to address these limitations by a complementary diary study where we delve deeper into the trends shown in the telemetry data set.

\subsection{Diary Study}
We complement our quantitative analysis with reports drawn from a diary study of employees from Microsoft reporting their experiences of remote meetings during COVID-19. Diary studies in HCI research \cite{rieman1993diary} have been particularly effective in capturing the nuanced experiences of information workers \cite{czerwinski2004diary,sellen1997paper}.

The diary study data collection ran between mid-April and mid-August 2020. Participants opted-in to the study from bulk email messages sent to internal mailing lists, with rolling recruitment between mid-April and mid-June 2020. Participants were asked to keep a guided diary for two months. Diary reminders were sent as a file via email to participants to set up in their calendars. The diaries consisted of a series of forms embedded in a secure company website. 24 total diary entries were requested, one entry approximately every two working days for two months total. The 24 guided entries were to be filled out in three cycles of eight topics: Physical workspace, Interaction, Productivity, Tools, Multitasking, Types of meetings, Time in meetings, and Approaches to meetings. The three diary entries on Multitasking used the following primary prompt: "What have you noticed about multitasking in recent online meetings?". This was followed by a list of sub-prompts: How and when you multitask; Why you multitask; Managing video and audio; Group expectations around multitasking; Impact of multitasking on productivity; Impact of multitasking on conversations; Features for multitasking; Impact of home life; Suggestions for improvement. Full methodological details are available in a technical report ~\cite{rintel2020methodology}.

849 total participants provided consent and were onboarded, of which 715 completed at least one diary entry. For those who filled out diaries: Gender coverage was 60\% Male, 39\% Female, and 1\% non-binary or preferred not to say. Age coverage was 5\% 18-24, 28\% 25-34, 30\% 35-44, 28\% 45-54, 8\% Above 55, and 1\% Prefer Not To Say. Roles covered were 41\% Business \& Sales, 30\% Development, 11\% Creative, Design, and UX, 11\% Technical Operations, 5\% Administration, and 2\% Research. Participants were drawn from almost all regions of the world, primarily 53\% North America, 20\% Europe, 12\% India, 4\% China, and 4\% South America.

Of 7045 total rows of diary responses, 819 responded specifically to the multitasking prompt\footnote{We did not include responses mentioning "multitasking" under other prompts.} from 413 unique participants. We randomly selected 100 responses (20 per month from April to August). To analyze the responses, we adopted the method of open coding~\cite{corbin2014basics}. Five researchers independently analyzed and coded the first 20\% of the interview transcriptions and met to discuss the codes until they were in complete agreement on the codes needed. Then, one researcher coded the remaining transcriptions but discussed any questions with the four other researchers so as to guarantee consensus on the codes. When these steps were finished, the whole research team met and thoroughly discussed the extracted content classification. Through sub-categorization and constant comparison~\cite{corbin1990grounded}, we consistently revised the emerging themes and the final themes presented in Sec 4, 5, 6 and 7 were developed. 



To ensure people's privacy in the diary study, we de-identified data before analysis (a participant key linked demographics to diary entries, and then all diary entries were scrubbed for names, places, and other identifying referents). Similar to the telemetry data set, only authorized researchers have access to the diary data.

\section{RQ0: Volume of Multitasking during Remote Meeting}
Our analysis suggests that multitasking during remote meetings is ubiquitous, and that people find themselves engaging in more multitasking during remote meetings compared to in-person meetings, possibly as a result of a shift in work rhythms, and the lower cost to get noticed.

\textbf{Multitasking intensity over time.}
In our telemetry dataset, from February to May, we find that 31.1\%, 30.9\%, 29.2\%, and 28.9\% meetings involve email multitasking, and 23.7\%, 23.1\%, 24.8\%, and 25.5\% meetings involve file multitasking. Putting these percentages in the societal context of the cognitive efforts and resources that information work costs \cite{spira2005cost}, our results suggest the importance of understanding and potentially intervening toward such behavior.

\textbf{More multitasking as a possible result of work rhythm adaptation.} Fig.~\ref{fig:workrhythm} illustrates the shift of work rhythms, characterized by the distribution of work related actions (emails sent, files edited and meetings attended) within a day, from Feb 2020 to May 2020. The email and file-related rhythms throughout the day remain roughly unchanged, indicating that people worked in a similar fashion on emails and files as they did in co-located settings. Meanwhile, there is a clear increase in the number of remote meetings, compared to pre-COVID-19 period -- note that the telemetry results do not necessarily support the conjecture that people are meeting more since not all in-person meeting frequencies were recorded in the telemetry. However, our diary study results suggest that people do perceive that they have more meetings, and this may be an important underlying cause of multitasking during remote meetings.

\begin{adjustwidth}{1em}{1em}
\emph{"I think this is more of a habit that developed now that folks don't have face to face meetings at all and that the number of meetings has increased so much that it is just more efficient to get the notes and reading out of the way during the meeting than work extra hours end of day or early next day to catch up"} (R498)
\end{adjustwidth}

\begin{adjustwidth}{1em}{1em}
\emph{"There are so many meetings that there is no time to look at email, or get work done in between. I try and work early in the morning and late at night, but as work flows in during the day and needs response, I find myself more and more just multitasking"} (R179)
\end{adjustwidth}

\begin{figure*}
  \centering
      \begin{subfigure}[Email Action Daily Rhythm]{
        \label{fig:Email_Day}
        \includegraphics[width=0.32\textwidth]{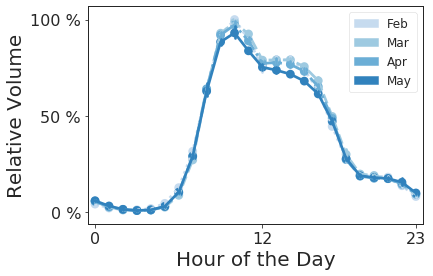}}
       \end{subfigure}
       \begin{subfigure} [File Action Daily Rhythm]{
        \label{fig:Meeting_Day}
        \includegraphics[width=0.32\textwidth]{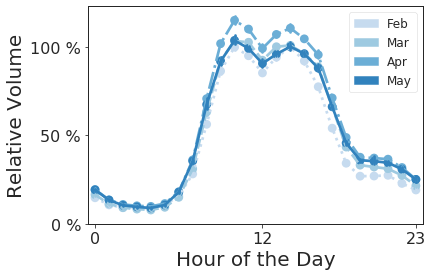}}
       \end{subfigure}
       \begin{subfigure} [Remote Meeting Daily Rhythm]{
        \label{fig:POI_revisit_T}
        \includegraphics[width=0.32\textwidth]{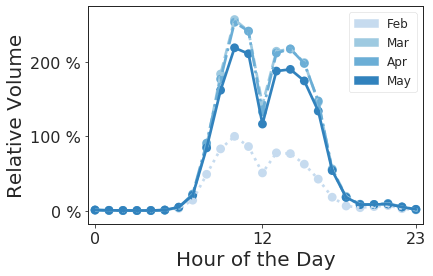}}
       \end{subfigure}
  \caption{Transition of worker work rhythms (i.e., distribution of work related actions over time within a day) from Feb. 2020 (pre-COVID-19) to May 2020 on email, file editing actions and remote meetings. While email and file usage remains stable, there is a clear increase in the volume (over twice as much) of remote meetings after the breakout of the pandemic. Results are normalized using maximum volume in Feburary.}
  \label{fig:workrhythm}
\end{figure*}




\textbf{Ease of turning off video and audio may encourage more multitasking.}
In comparison with traditional face-to-face meetings, remote meetings make it much easier for people to stay in the background by turning video off/muting themselves. Given that multitasking has been culturally associated with impoliteness \cite{przybylski2013can}, we assume more multitasking during remote meetings may be caused by the lower probability of getting noticed by others when multitasking. In our diary study responses, we found that turning off the video camera or muting the microphone was closely related to multitasking behavior, as mentioned by many (32\% of responses).

\begin{adjustwidth}{1em}{1em}
\emph{"I typically will not multi-task if I have my video on, because people can definitely tell when you're not paying attention. Sometimes I will choose to turn my video on purely so that I am not tempted to multi-task. If I am an optional participant in a meeting, or I am just listening in, unsure if the agenda really calls for my participation, then I am more likely to keep my video off and openly multi-task until someone says my name."} (R10)
\end{adjustwidth}

\begin{adjustwidth}{1em}{1em}
\emph{"In general, I have a feeling that in our group the expectation is that participants do not multitask during meetings, but who knows what you are doing if the camera is off. (So yeah, that's when I turn off my camera too.)"} (R14)
\end{adjustwidth}
\section{RQ1: WHAT FACTORS ARE ASSOCIATED WITH MULTITASKING}

\begin{figure*}[tb]
       \begin{subfigure} [Meeting duration, >80 mins ($P<0.001$), 40-80 mins ($P<0.001$), 20-40 mins ($P<0.001$)]{
        \label{fig:EmailRegression_duration}
        \includegraphics[width=0.32\textwidth]{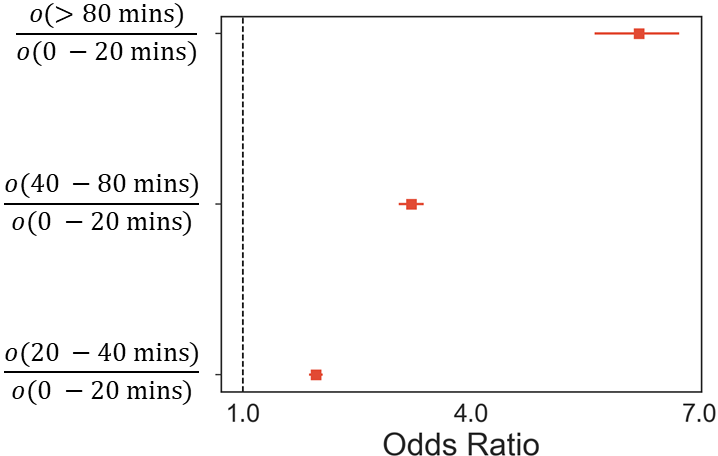}}
       \end{subfigure}
       \begin{subfigure} [Meeting size, >10 attendees ($P=0.021$), 6-10 attendees ($P<0.001$), 4-5 attendees ($P<0.001$), 3 attendees ($P=0.021$)]{
        \label{fig:EmailRegression_size}
        \includegraphics[width=0.34\textwidth]{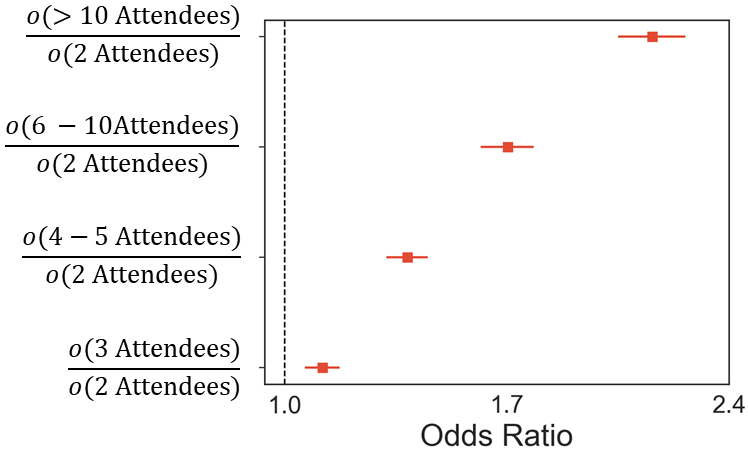}}
        \end{subfigure}
       \begin{subfigure} [Meeting type, scheduled ($P=0.012$), recurring ($P<0.001$), broadcast ($P=0.880$) ]{
        \label{fig:EmailRegression_type}
        \includegraphics[width=0.3\textwidth]{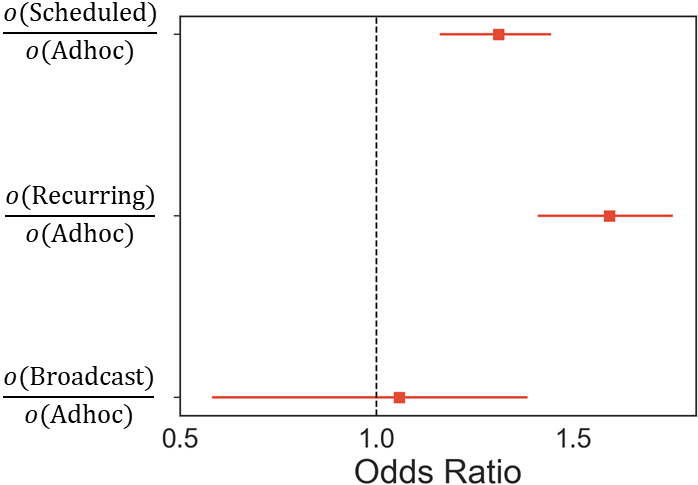}}
       \end{subfigure}
       \begin{subfigure} [Hour of the day, morning ($P<0.001$), afternoon ($P<0.001$)]{
        \label{fig:EmailRegression_hour}
        \includegraphics[width=0.32\textwidth]{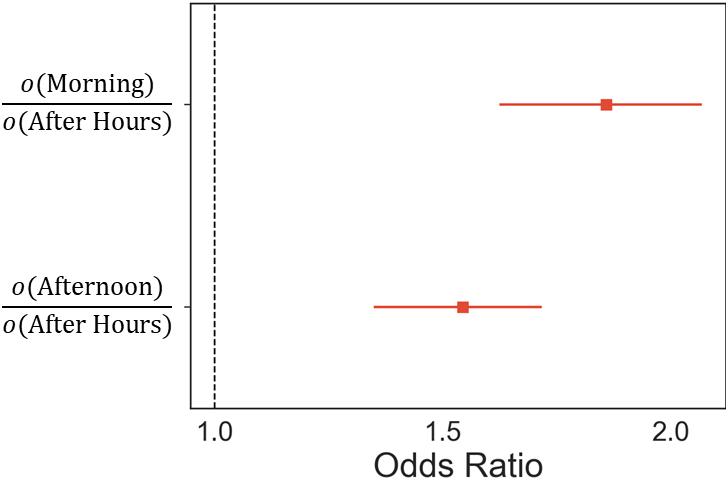}}
       \end{subfigure}
      \begin{subfigure}[Day of the Week, Thursday ($P=0.003$), Wednesday ($P=0.003$), Tuesday ($P<0.001$), Monday ($P=0.001$)]{
        \label{fig:EmailRegression_Weekday}
        \includegraphics[width=0.32\textwidth]{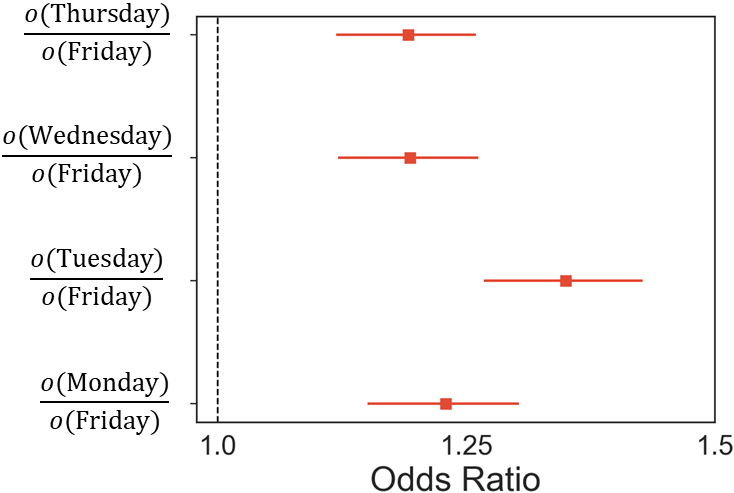}}
       \end{subfigure}
 \caption{Conditional logistic regression results on the relationship between email multitasking and remote meeting characteristics. We find significant associations between email multitasking and meeting duration, meeting size, meeting types, hour of the day and day of the week.}
 \label{fig:Regression_email}
\end{figure*}

\begin{figure*}[tb]
       \begin{subfigure} [Meeting duration]{
        \label{fig:EmailMultitasking_duration}
        \includegraphics[width=0.3\textwidth]{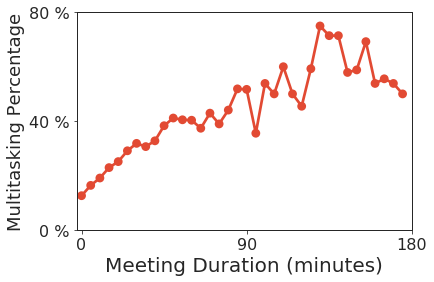}}
       \end{subfigure}
       \begin{subfigure} [Meeting size]{
        \label{fig:EmailMultitasking_size}
        \includegraphics[width=0.3\textwidth]{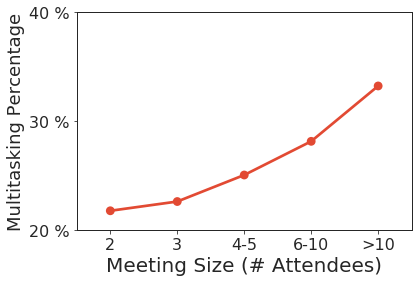}}
       \begin{subfigure} [Meeting types]{
        \label{fig:EmailMultitasking_type}
        \includegraphics[width=0.3\textwidth]{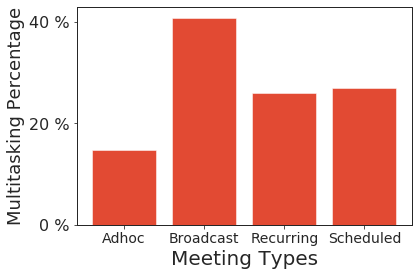}}
       \end{subfigure}
       \end{subfigure}
       \begin{subfigure} [Hour of the day]{
        \label{fig:EmailMultitasking_hour}
        \includegraphics[width=0.3\textwidth]{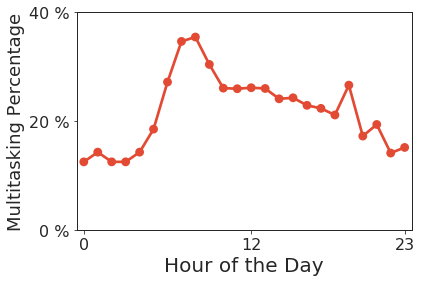}}
       \end{subfigure}
      \begin{subfigure}[Day of the Week]{
        \label{fig:EmailMultitasking_day}
        \includegraphics[width=0.3\textwidth]{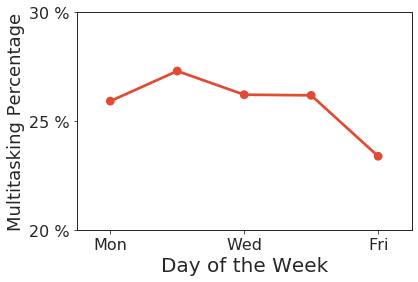}}
       \end{subfigure}
 \caption{Proportion of user-meeting pairs with email multitasking versus meeting characteristics measured by telemetry data. People multitask more in longer meetings, larger meetings, recurring/scheduled meetings, morning meetings and Tuesdays.}
 \label{fig:Email_Multitasking}
\end{figure*}

Our analysis suggests that both intrinsic meeting characteristics and external factors are correlated with remote meeting multitasking, as discussed in detail below.

\subsection{Intrinsic Meeting Characteristics}
\textbf{More multitasking happens in large meetings.} As shown in Fig.~\ref{fig:EmailRegression_size} and Fig.~\ref{fig:EmailMultitasking_size}, larger meetings generally involve more multitasking. The odds of email multitasking in 3 attendee meetings, 4-5 attendee meetings, 6-10 attendee meetings, and >10 attendee meetings are 1.12 ($P=0.021$), 1.39 ($P<0.001$), 1.70 ($P<0.001$) and 2.16 ($P<0.001$) times the odds of the meetings with only one or two attendees. This could be explained by the fact that participants need to more actively focus on the meeting conversations when the meetings are small. Our empirical finding is also supported by evidence from the diary study, e.g., 

\begin{adjustwidth}{1em}{1em}
\emph{"If it's a large audience, like a webcast or an internal session on some tech topic, I do multitask more."} (R4)
\end{adjustwidth}

\begin{adjustwidth}{1em}{1em}
\emph{"In the big meetings, like town halls, I tend to stop and actually listen when something of interest is being said. the rest of the time, I seem to not focus on work at all. In small meetings, I generally don't multitask at all anymore. It takes all of my brainpower to focus on the conversation."} (R182)
\end{adjustwidth}

\textbf{More multitasking happens in long meetings.}
We also observe that more multitasking happens in longer meetings. From telemetry data, we observe that the odds of email multitasking in 20-40 minute meetings, 40-80 minute meetings, and >80 minute meetings are 1.96, 3.22 and 6.21 times the odds of 0-20 minutes meetings ($P<0.001$), as illustrated in Fig.~\ref{fig:EmailMultitasking_duration} and Fig.~\ref{fig:EmailRegression_duration}). 
As supported by diary study responses, many people mention that they simply cannot concentrate for a long time and thus engage in multitasking during long meetings.

\begin{adjustwidth}{1em}{1em}
\emph{"Additionally, meetings seem to be longer (e.g., I have a three hour brainstorming session with my team today) and I cannot focus on the meeting that long alone. Then I also tend to work on other tasks from now and then."} (R21)
\end{adjustwidth}

\textbf{Morning meetings involve more multitasking.}
The time schedule of the meeting also plays an important role in multitasking behavior. Our email multitasking data analysis suggests that morning meetings involve more multitasking compared to afternoon and after hours: the odds of multitasking in the morning are 1.86 times the odds of the after hour meeting baseline ($P<0.001$), and the odds of email multitasking in the afternoon are 1.54 times the odds of the after hour meeting baseline ($P<0.001$). We argue that this observation may be closely related to the daily work rhythms of individuals. As demonstrated by prior work \cite{mark2014bored}, in the afternoon people are generally more focused. We also find supporting evidence in the diary study results (6\% of responses)

\begin{adjustwidth}{1em}{1em}
\emph{"I will try to glimpse at email more if a meeting is the first thing I do for work in the morning - so that's schedule-related."} (R621)
\end{adjustwidth}

\textbf{More multitasking happens in recurring and scheduled meetings compared to ad hoc meetings.}
Next, our results suggest that multitasking is more likely to happen in recurring and scheduled meetings compared to ad hoc meetings. We find significant associations between email multitasking and meeting types in our telemetry data. Specifically, the odds of email multitasking in recurring and scheduled meetings is 1.59 ($P<0.001$) and 1.31 ($P=0.012$) times the odds of multitasking in ad hoc meetings. Ad hoc meetings generally involve a specific focus relevant to the specific attendees, while scheduled meetings, especially recurring meetings, are more likely to involve broader information sharing which does apply equally to each attendee.
\begin{adjustwidth}{1em}{1em}
\emph{"I just came off an online call with my larger design sync, it's a 30 min recurring meeting where we share topics as our design teams are in [two cities]. I didn't need to present, I was a listener, so I found myself responding to Teams messages, emails etc as the call was going on."} (R42)
\end{adjustwidth}

\textbf{More multitasking happens Monday through Thursday compared to Friday.}
The telemetry data also demonstrated that Friday has a relatively lower likelihood of multitasking, compared to other days of the week. The odds of email multitasking on Tuesday are 1.35 times the odds of multitasking on Friday ($P<0.001$), followed by Monday (1.23 times, $P=0.001$), Wednesday (1.19 times, $P=0.003$) and Thursday (1.19 times, $P=0.003$). While this pattern corroborates findings from prior work ~\cite{mark2014bored}, we note that the result might not generalize broadly, especially for Fridays, since Microsoft encourages fewer meetings on Friday, so the findings might be company-specific.

\textbf{More multitasking in meetings of lower relevance and engagement.}
People also frequently mentioned (17 \% responses) that they multitask during meetings they find irrelevant or have lack of interest or engagement in, which might be the underlying reason why people multitask more in larger group sizes and longer meetings.


\begin{adjustwidth}{1em}{1em}
\emph{"I tend to multitask more in larger group meetings online. Larger meetings often have topics on the agenda not directly related to what I'm working on day to day so my mind tends to wander when the topic of discussion is irrelevant. I'm definitely aware of trappings of a divided mind. I'm not necessarily productive on these other tasks. So I normally engage in menial work like cleaning and organizing files."} (R11)
\end{adjustwidth}

\begin{adjustwidth}{1em}{1em}
\emph{"I myself frequently have a web page, source code, or build window open in another window, and I divide my attention - most often when the meeting goes to topics that don't concern me, as most of my meetings do."} (R15)
\end{adjustwidth}

\begin{adjustwidth}{1em}{1em}
\emph{"When meetings have a lot of topics or don't apply to me, I start multitasking."} (R16)
\end{adjustwidth}



Sometimes people lose their concentration due to high cognitive load under such meetings of low relevance,

\begin{adjustwidth}{1em}{1em}
\emph{"I've noticed that I only multitask when I am tired and it is difficult for me to focus on the ongoing meeting. And I don't do difficult tasks either, at most I am checking if my PR went through or kick off a build or just look at pictures of cats. "} (R14)
\end{adjustwidth}

\begin{adjustwidth}{1em}{1em}
\emph{"It's really hard to focus, I don't know what people are trying to say or what the actions items after or why we're discussing certain items. When I can't focus or understand what's going on, I tend to check out and look at other things or do something where I feel engaged and useful. "} (R346)
\end{adjustwidth}

\subsection{Extrinsic Factors}
\textbf{People multitask during meetings to catch up on other work.}
Another major reason (39 \% responses) why people multitask is to catch up on their work. Given the increasing number of meetings compared to the in-person work experience, people find they are having a hard time completing all of their work in time.

\begin{adjustwidth}{1em}{1em}
\emph{"It needs to happen or you cant get all your work done "} (R167)
\end{adjustwidth}

\begin{adjustwidth}{1em}{1em}
\emph{"But these days I am having a lot of meetings, making it hard to find time to get work done. So, feel super tempted to multi task, if not entire day goes in meetings before real work gets started. Another reason to multitask is deadlines... "} (R12)
\end{adjustwidth}

\begin{adjustwidth}{1em}{1em}
\emph{"My team is often quite bad at sticking to an agenda, so I find myself multitasking as a way to feel like I'm still able to "get things done"while I'm 'sitting in' on these meetings.  As a designer, that often means that I'm clicking around in Figma.  Increasingly, I am also trying to use meeting time to catch up on context for other meetings.  This makes paying attention in any meeting very difficult, but with the volume of meetings and the complexity of the context I feel I need to maintain, I often feel like I have no choice."} (R9)
\end{adjustwidth}

\begin{adjustwidth}{1em}{1em}
\emph{"Lately (since COVID) I've been forced to multitask during meetings to meet the deadlines that I've been given (and even then, I still don't always make it)... I don't have enough hours in the day to do all of the work that is required of me."} (R346)
\end{adjustwidth}

\textbf{People multitask during meetings due to external distractions.}
We also find people frequently multitask during remote meetings as a result of external distractions - under such situations, people do not purposely multitask, but their attention gets attracted by external sources. Two major classes of distractions are interface design, and the home working environment. As collaboration moves online in remote work, people are interacting with digital tools more than they used to when co-located at work, and people are mentioning that interface designs can be the cause of multitasking behavior, especially pop-ups, e.g.,

\begin{adjustwidth}{1em}{1em}
\emph{"I multi-task in almost every meeting that isn't 1-1 and even in 1-1 meetings it's hard not to multi-task because email and teams chats are popping in.  In person for 1-1 i would lock my computer and focus entirely on the person and this is super hard in the WFH setup."} (R502) 
\end{adjustwidth}

\begin{adjustwidth}{1em}{1em}
\emph{"There are a lot of people are multitasking as we are using Teams. Teams is prompting up that someone is trying to get hold of us. This lures us in to checking quickly in to who it is an what they want"} (R664)
\end{adjustwidth}

On the other hand, as pointed out by prior qualitative work~\cite{allen2015effective}, remote work involves more distractions from the home working environment that could lead people to multitasking, e.g.,
\begin{adjustwidth}{1em}{1em}
\emph{"Since COVID19 the multitasking also includes:  - answering children's questions  - preparing food for children  - Helping children with school work  - resolving children's disagreements "} (R183)
\end{adjustwidth}

\textbf{People multitask during meetings for anxiety relief.}
Finally, some participants mentioned that anxiety over the COVID-19 pandemic lead them to seek methods for maintaining focus, such as conducting a low cognitive effort non-work activities while monitoring meetings.
\begin{adjustwidth}{1em}{1em}
\emph{"The current situation has increased my general anxiety. This means I have more difficulty focusing on tasks - including meetings. I have been multi-tasking with non-work tasks (i.e. a colouring game on my phone!) quite a bit as I find this actually enables me to focus better on the meeting."} (R5)
\end{adjustwidth}

\begin{adjustwidth}{1em}{1em}
\emph{"Doing some exercises with my shoulders/back during meetings (with video turned off) is great."} (R13)
\end{adjustwidth}

\section{RQ2: WHAT DO PEOPLE DO WHEN MULTITASKING}
Our analysis further shows that people engage in work-related and non-work-related tasks when multitasking during remote meetings.

\subsection{Work-related tasks}
\label{sec:file_multitasking_qualitative}

Communication with coworkers is one of the most frequently mentioned multitasking behavior, since people generally consider that communication is quick to complete without much need to focus, or "light weight''. In fact, 29 \% of diary study participants mention that they engage in email multitasking, which further confirms and strengthens our motivation to analyze email multitasking using the telemetry data set.

\begin{adjustwidth}{1em}{1em}
\emph{"I might use that time to reply to simple emails that don't require much thinking (so I can also pay attention to the meeting)."} (R170)
\end{adjustwidth}

\begin{adjustwidth}{1em}{1em}
\emph{"And there are different levels of multitasking, meaning the kinds of multitasking things I will do during meetings can range from writing down a quick reminder to myself on a different topic (takes a few seconds) to triaging email which takes very short attention shifts and I can come back into the proceedings easily, to more cognitively demanding tasks like writing emails, responding to IM threads etc." } (R171)
\end{adjustwidth}

Meanwhile, several diary study participants mentioned that they also frequently check and test scripts that take time to run during remote meetings, which is also rather lightweight. 

\begin{adjustwidth}{1em}{1em}
\emph{"There have been situations where I've multitasked in the sense of responding to email or checking on the results of a long running job while in a meeting that is more focused on consumption of information rather than on my own contributions."} (R180)
\end{adjustwidth}

\begin{adjustwidth}{1em}{1em}
\emph{"Because I have done a lot testing and building scripts. It takes time, so I can work with other stuff while waiting for the result."} (R337)
\end{adjustwidth}

People also mentioned that they engaged in file multitasking, yet these activities are often related to the meeting, e.g., notes taking (R182, R183, R344), checking relevant files (e.g. \emph{"in a meeting discussing aspects of the project I linked to the latest documentation."}, R174), etc., which could be a possible explanation of why there is no significant relationship between file-related multitasking and meeting characteristics. 


\subsection{Non-work-related tasks}

Non-work tasks is also an emerging theme in the diary study responses. For instance, checking social media as a break from work.

\begin{adjustwidth}{1em}{1em}
\emph{"I've been multitasking both for personal and professional things - answering emails and chats for work, taking social media and texting breaks for personal."} (R168)
\end{adjustwidth}

Given that working from home environment under COVID-19, people also reported that they engaged in eating, exercise (primarily for anxiety relief and wellness) and other household chores.

\begin{adjustwidth}{1em}{1em}
\emph{"As my kitchen is very close to the desk with my computer, I also get food or drinks from there during the meetings more often."} (R20)
\end{adjustwidth}

\begin{adjustwidth}{1em}{1em}
\emph{"For me personally, I am more likely to do another task that does not require the computer like managing my to do list, writing down notes, cleaning up my office and desk, eating etc."} (R171)
\end{adjustwidth}

\begin{adjustwidth}{1em}{1em}
\emph{"It has been beneficial to walk around my house while on meetings (that don't require me to be in front of my computer).  I'm able to put dinner in the oven, feed the dogs, etc."} (R668)
\end{adjustwidth}

\section{RQ3: THE CONSEQUENCES OF MULTITASKING}

Finally, we present our findings on the consequences of multitasking during remote meetings. Although multitasking is typically associated with negative outcomes such as decreased task performance ~\cite{monk-interruptionCost, cutrell-interruptionCost}, difficulties in decision making \cite{speier-decisionmakingCost} and negative affect \cite{zijlstra-affect, Bailey2006OnTN}, our participates report that in-meeting multitasking leads to both positive and negative outcomes.

\subsection{Positive Outcomes}
\textbf{Multitasking may help boost productivity.} First, multiple participants (15 \%) mentioned that multitasking helps boost their productivity, which echos prior works on the benefits of multitasking on efficiency \cite{dabbish-interruptionBenefits, maglio-peripheral, rich-multitaskingBenefits}. Given that, under current remote work settings, there are many more remote meetings compared to the pre-COVID-19 period, people explained how multitasking helps them to get work done. Here's a representative response, 

\begin{adjustwidth}{1em}{1em}
\emph{"There are some benefits to multi-tasking. I've been able to get more work done. I've been less frustrated by meetings that weren't very useful to me. I haven't made any significant mess ups in a meeting yet when I've done it."} (R330)
\end{adjustwidth}
Multitasking is at its most productive when workers understand that their own and others' attention in a meeting is a spectrum about which they can make active meeting choices \cite{kuzminykh2020low}.
\begin{adjustwidth}{1em}{1em}
\emph{"I find myself multi-tasking in meetings that do not require my attention, but not in meetings that do. In some ways, I may be more productive given the ease of multi-tasking in remote meetings."} (R508)
\end{adjustwidth}

People also mentioned the positive experience with meeting-related multitasking behavior, such as note taking and searching for information, e.g.,

\begin{adjustwidth}{1em}{1em}
\emph{"The type of multitasking that I feel it has a positive impact on my productivity is when I am taking notes during a meeting or when I am navigating the internet to find some relevant information that is being discuss in the meeting."} (R668)
\end{adjustwidth}




\subsection{Negative Outcomes}
\textbf{Multitasking leads to loss of attention/engagement.}
Nevertheless, remote meeting multitasking does cause negative consequences. Among them, the most frequently (36 \%) mentioned negative aspect is loss of attention/engagement, where people lose track of the meeting content (which sometimes is important) due to multitasking activities, as demonstrated by the following cases, 

\begin{adjustwidth}{1em}{1em}
\emph{"Its easy to get distracted by multitasking and miss something in the meeting."} (R2)
\end{adjustwidth}

\begin{adjustwidth}{1em}{1em}
\emph{"I have to channel my concentration on 1 primary task -- whether that be the meeting itself or the side work I'm doing while listening in."} (R346)
\end{adjustwidth}

\begin{adjustwidth}{1em}{1em}
\emph{"If you leave the meeting window to open a deck or other files, it's hard to get back to the meeting window. Same with chats - if you leave the meeting window to send chats to other people, it's hard to get back to the meeting."} (R344)
\end{adjustwidth}

These observations are well-aligned with previous conclusions on the impact of multitasking behavior on people's attention and task resumption \cite{Bailey2006OnTN, altmannResumption}.

\textbf{Multitasking leads to mental fatigue.} Moreover, we find that remote meeting multitasking behavior does have an impact on well-being: some participants reported that they feel tired after multitasking during remote meetings.

\begin{adjustwidth}{1em}{1em}
\emph{"I tire a bit more with so many meetings and multitasking."} (R72)
\end{adjustwidth}

\textbf{Multitasking can be disrespectful.} One final downside of multitasking is that it has been sometimes regarded as an inappropriate behavior during remote meetings. Some participants explicitly associate multitasking with being impolite.

\begin{adjustwidth}{1em}{1em}
\emph{"I tend to do it less while on video, so they can't tell I'm being rude."} (R173)
\end{adjustwidth}

\begin{adjustwidth}{1em}{1em}
\emph{"I rarely multitask when on camera--it seems rude."} (R673)
\end{adjustwidth}

\begin{adjustwidth}{1em}{1em}
\emph{"People are becoming a bit more brazen; it's sometimes unbelievably clear that they aren't paying close attention to the discussion and with the current WFH situation it's hard not to notice eyes straying, backlight (windows) changing, etc. It adds another dimension of rudeness but I also do it to others, sometimes not realizing when I follow the stray notification."} (R343)
\end{adjustwidth}

\begin{adjustwidth}{1em}{1em}
\emph{“I have gotten caught off guard a few times though. Someone will ask me a question and I'll  have to ask them to repeat it. They don't seem upset, but I'm still embarrassed/ashamed when it happens. This is mostly due to the fact that I know I have slightly negative feelings toward other people when they get caught in the same situation, or when they ask a question that had been asked and answered recently. I don't want to be one of those people or be thought of as one of those people who is not paying attention."} (R330)
\end{adjustwidth}
\section{Best-practice guidelines for remote meetings}
\label{sec:best_practice}


Remote meetings have become the primary way that people connect and collaborate while working from home. As the number and duration of remote meetings has increased, people appear to have been left with less time to focus on their work and thus have gotten into the habit of multitasking to catch up. This draws participants' attention away from the meeting, and can lead to mental fatigue and disrespectful behaviors. Based on our findings, we propose several practical remote meeting best practices that meeting organizers can use to help people attending the meeting actually attend to the meeting. We note, however, the importance to consider specific worker and corporation contexts when applying these guidelines.

\textbf{Avoid important meetings in the morning.}
As demonstrated by Fig. \ref{fig:workrhythm}, people still adhere to a similar ``double peak'' daily work rhythm~\cite{mcduff2019longitudinal} as they did in pre-COVID, co-located work settings. Through our regression analysis, we find that email multitasking behavior occurs most often in the morning, which coincides with the fact that email actions peak in the morning. Prior evidence also suggests that people are most focused in mid afternoons~\cite{mark2014bored}. Therefore, our results suggest that meeting organizers might avoid scheduling important meetings in the morning, when it is harder for people to concentrate. However, multitasking in the morning is not always bad, as confirmed by our diary study participants (e.g., “balancing home and work”). In fact, scheduling light-weight meetings in the morning may actually help people smoothly transition from ``home'' to ``work'' mode under remote work settings \cite{williams2018supporting,jachimowicz2020between}.

\textbf{Reduce the number of unnecessary meetings.}
Many participants stated that they multitasked because they found that the content of some meetings did not apply to them, especially for information sharing and daily stand up meetings. In the telemetry data analysis, we also found that there was much more email multitasking in recurring and scheduled meetings compared to ad hoc meetings. Given there are so many meetings people now need to attend, meeting organizers should reconsider the necessity of numerous meetings, or the frequency of such meetings, so as to help people better focus. The organizers may also consider sharing information asynchronously. For example, sending out a recording of a presentation for attendees to watch on their own, and then only use the meeting for discussion.





\textbf{Shorten meeting duration and insert breaks.}
In our telemetry analysis, we found that longer meetings are associated with more multitasking behavior, which is also verified by the qualitative evidence. As suggested by prior literature, humans have an upper time limit where they can fully engage and pay attention~\cite{mark2017blocking}, thus we suggest that meeting organizers should shorten the duration of meetings, or insert breaks when meetings have to run long.

\textbf{Encourage active contribution from the appropriate number of attendees.}
Finally, many participants mentioned that they multitasked because they don't have anything to actively contribute to the meeting discussion. They generally muted themselves and turned off their video in such scenarios. If organizers want the full attention of participants in an important meeting, they should encourage participants to actively engage through stimulating interactions, especially if it is a large meeting with a variety of attendees. For meetings where active engagement is particularly important, then the invitee list should be as small as necessary to achieve the right level of engagement from attendees. 

\textbf{Allow space for positive multitasking.} 
Our findings suggest that multitasking can be positive under remote environment. Therefore, meeting organizers could consider creating personalized meeting agenda so that people are aware of the timing when relevant agenda items come up. Organizers can also implement a convention where video-on implies full attention, and video-off signals multitasking.


\section{Design implications for better support of remote meetings}

We have seen that multitasking in remote meetings is a complex behavior, with both positive and negative aspects. It can help people be more productive, but may also reduce attention, increase fatigue, and appear impolite. While culturally the term multitasking may have a negative connotation \cite{przybylski2013can}, our study adds to a growing body of work that reevaluates differential attention and multitasking in remote work contexts \cite{kuzminykh2020low} and for users of all abilities \cite{das_towards_frth}. Based on our findings, we argue that given the complext nature of remote meeting multitasking, it is important to encourage its positive aspects while reducing its negative implications. In this section we discuss several ways productivity tools might do this.

\textbf{Support a `focus mode' for remote meetings.}
Our analysis shows that pop-ups in current software interfaces during remote meetings distract people from the meeting itself. To alleviate such distractions, we envision future collaboration platforms having a remote meeting `focus mode'. After people choose to enter the mode 
(e.g., for a very important meeting), the tool could block all standard pop-up messages, emails, etc., so as to help them concentrate on the meeting itself. The focus mode could also employ a multitasking alert feature. People could give permission to the app to track their behavior in other applications or even their other devices. For non-meeting focused behavior  (speaking, screensharing, in meeting parallel chat, etc.), the app could alert people about their multitasking and reflect back the reason (e.g., not being able to absorb all of the information within the meeting). As such, the feature could help people engage in the meeting and avoid unintended loss of attention.

\textbf{Support other types of engagement during remote meetings.}
Some positive multitasking fits the meeting purpose, but technically requires moving outside the meeting window which may increase the risk of being distracted from the meeting. For instance, people often need to work outside the actual meeting window or platform to take notes or work on files, switching back and forth between this work and the meeting. The ability to have more windows, split views, or even a more broadly-defined meeting space could not only reduce the potential for distraction but also improve the shared use of these resources. Given the various reasons for remote meeting multitasking, it is also important to not arbitrarily consider it unacceptable. Meetings could enable attendance along a scale of high to low engagement, to help set attendees' expectations and more closely match actual behaviour~\cite{kuzminykh2020low}. 

\textbf{Help people decide which meetings to attend.}
Our study suggests that over-multitasking or negative multitasking is associated with the increasing number of meetings during remote work. Apart from organizational leaders actively reducing the number of meetings, future remote meeting tools could develop a feature for people to self-rank the importance of each meeting, or recommend an importance level assigned to each meeting for each individual based on meeting characteristics (e.g., content, size, attendees, etc.), and add the importance level of each meeting on the person's calendar apps and video conferencing platforms. As such, people will have a better idea for each day which meeting is important and when they should pay special attention in order to not miss key information (thus avoiding multitasking). For less important meetings, the system could recommend ways to catch-up later, alternative ways to attend, or help notify the meeting organizer. 

\textbf{Help people skip some parts of the meeting.}
As suggested by our findings, even within the same meeting, certain parts of the meeting may not be so important as other parts for a specific attendee: the attendee may only be interested in a particular section, and works on other things except for that section. If meetings have agendas, one solution would be to for organizers and attendees to flag expected attention per item, and add increasing and decreasing visible attention when relevant. 
Future tools can also help better support personalized meeting content importance tracking for attendees. For instance, the system could make use of real time transcriptions of meeting and compare the topic similarity to those that the attendee might be interested in.

\section{Limitations and generalizability}
While our findings build on rich telemetry and diary study data to extend what was previously know about multitasking in remote meetings, it is important to consider them in the context of several limitations.
For one, the data are drawn from one global information technology company (and the telemetry data focus on US workers only), with most participants being information workers. As a result, our findings may not generalize to other worker types under remote settings, or to other cultures.  Further, the analyzed data were collected during the COVID-19 pandemic era, but we were not able to distinguish remote work effects from COVID-19 effects (e.g., the impact of remote working from home on people's mental wellbeing) on multitasking behavior. Additionally, while telemetry data is valuable in providing a large-scale, realistic picture of behavior, it does not provide the motivation motivating that behavior \cite{dumais2014understanding}, we are not able to distinguish whether the email and file actions we observed were related to the meeting or not. It is likely that some of the positive multitasking observations may be false positives. Finally, the current diary study does not cover every aspect with regard to remote meeting multitasking, which can be addressed in future work,  e.g.,  references to side-channels, such as Whatsapp, Facebook or other apps that cannot be instrumented through telemetry data analysis and people's perception of multitasking. Nevertheless, we believe that  this work presents the most comprehensive analysis of remote meeting multitasking behavior currently available, and could extend to scenarios beyond the narrow-sense of workspace meetings (e.g., real-time distance education that leverages remote meeting tools \cite{chen2020large}). 
We leave the above mentioned limitations as future work.
\section{Conclusion}
In this paper, we presented a large-scale and mixed-methods study of multitasking behavior during remote meetings. We analyzed a large-scale telemetry dataset and conducted a longitudinal diary study during COVID-19 period. Our analysis leads to practical guidelines for remote meeting attendees and design implications for productivity tools, both of which can improve remote meeting experiences. Our results point to the importance of multitasking in the consideration of remote meetings, both with respect to their social and technical components. On the one hand, how remote meetings are scheduled and structured are significantly associated with when and to what extent people divide their attentions. On the other hand, multitasking could imply both positive and negative outcomes for individual worker and work group. More future research efforts are needed to address remote meeting experience as it becomes mainstream, and our work provides a foundational and timely understanding of such.


\bibliographystyle{ACM-Reference-Format}
\bibliography{other_ref,ref}

\appendix
\section{Alternative Regression Analysis}
\label{sec:supp}

To test the robustness of our regression model (Section~\ref{sec:regression_analysis}), we conducted an alternative analysis using Generalized Linear Mixed Effects Models~\cite{seabold2010statsmodels}. Specifically, we estimated random intercepts for workers and approximated the posterior using variational Bayes. The alternative results are shown in Table.~\ref{tbl:alternative_regression}, and they are qualitatively similar to those we present in our main analysis.

\begin{table}[!htbp]
\begin{tabular}{lcc} \hline
Effect                  & Post. Mean & Post. SD \\ \hline
Day:Monday                  &  0.2345    &  0.0308    \\
Day:Tuesday                 &  0.3154    &  0.0277  \\
Day:Wednesday               &  0.2052   &  0.0276  \\
Day:Thursday                &  0.1940   &   0.0287   \\
Type:broadcast               &  0.8271    &  0.1710   \\
Type:recurring               &  0.5125    &  0.0191   \\
Type:scheduled               &  0.4235    &  0.0199   \\
Size:3               &  0.0591    &   0.0335  \\
Size:4-5                &  0.2152   &  0.0293  \\
Size:6-10               &  0.3656   &  0.0311  \\
Size:\textgreater 10    &  0.5788   &  0.0288   \\
Hour:morning                 &   0.6053    &  0.0186    \\
Hour:afternoon               &   0.3760    &  0.0202  \\
Duration:20-40 min           &  0.7863    &  0.0203   \\
Duration:40-80 min           &  1.3049    &  0.0265   \\
Duration:\textgreater 80 min &  1.9447    &  0.0699  \\ \hline
\end{tabular}
\caption{Alternative regression results with Generalized Linear Mixed Effects Models~\cite{seabold2010statsmodels}. The model includes random intercepts for workers and is approximated using variational Bayes. The results are qualitatively similar to those of our main analysis. \label{tbl:alternative_regression}}
\end{table}

\end{document}